\documentclass[12pt]{iopart}

\usepackage{graphicx}
\usepackage{iopams}
\begin{document}

\title{Upward curvature of the upper critical field and the V-shaped pressure dependence of $T_c$ in the noncentrosymmetric superconductor PbTaSe$_2$}

\author{J. R. Wang$^1$, Xiaofeng Xu$^{1}$ \footnote[1]{Electronic address: xiaofeng.xu@hznu.edu.cn}, N. Zhou$^1$, L. Li$^1$, X. Z. Cao$^1$, J. H. Yang$^1$, Y. K. Li$^1$, C. Cao$^1$, Jianhui Dai$^1$, J. L. Zhang$^2$, Z. X. Shi$^3$, B. Chen$^4$, Zhihua Yang$^5$}
\address{$^{1}$Department of Physics and Hangzhou Key Laboratory of Quantum Matters, Hangzhou Normal University, Hangzhou 310036, China\\
$^{2}$High Magnetic Field Laboratory, Chinese Academy of Sciences, Hefei 230031, China\\
$^{3}$Department of Physics, Southeast University, Nanjing 211189, China\\
$^{4}$Department of Physics, University of Shanghai for Science $\&$ Tehcnology , Shanghai, China\\
$^{5}$Xinjiang Technical Institute of Physics and Chemistry of Chinese Academy of Sciences, 40-1 South Beijing Road, Urumqi 830011,
China\\}

\date{\today}

\begin{abstract}
The temperature evolution of the upper critical field $H_{c2}(T)$ in the noncentrosymmetric superconductor PbTaSe$_2$ was determined via
resistivity measurements down to 0.5 K. A pronounced positive curvature in the $H_{c2}$-$T$ phase diagram was observed in the whole
temperature range below $T_c$. The Seebeck coefficient $S(T)$ in the temperature range 5K$\leq$$T$$\leq$350K was found to be negative in
sign, modest in magnitude and non-linear in temperature. In addition, the superconducting transition temperature $T_c$ under hydrostatic
pressure shows a marked non-monotonic variation, decreasing initially with the applied pressure up to $P_c$$\sim$5-10 kbar but then rising
with further pressurization. The underlying physical mechanisms of all these findings have been discussed.
\end{abstract}


\maketitle

\section{Introduction}

Superconductors lacking inversion symmetry have attracted tremendous research interest since the discovery of superconductivity in the
noncentrosymmetric (NCS) CePt$_3$Si \cite{CePtSi}. In the presence of inversion symmetry, Cooper pairs are either spin-singlet or
spin-triplet, i.e, having well-defined parity. However, in NCS superconductors, the link between spatial symmetry and the Cooper pair spins
may be broken and the admixture of singlet-triplet pairing is possible \cite{Yuan06,Yuan13,Tafti13-LuPtBi}. This is because the
antisymmetric spin-orbit coupling (SOC) in the NCS superconductors may split the Fermi surface into two sheets with distinct spin
helicities \cite{Yuan06,Shimozawa14}. The splitting of the Fermi surface therefore favors both the intra- and inter-band pairing, resulting
in the admixture of spin-singlet and spin-triplet components, with the ratio of these two being tunable by the strength of SOC
\cite{Yuan06,Zheng07}. Recently, NCS superconductors with \textit{strong} SOC were proposed as potential platforms to realize so-called
topological superconductivity \cite{ZhangSC10,Lu10,Hasan10}. The combination of strong SOC and broken inversion symmetry can give rise to
topological superconductivity of class DIII \cite{Hasan10}.

The NCS compound PbTaSe$_2$ was recently reported to be superconducting below $T_c$$\sim$3.7 K \cite{Cava14}. The crystal structure of
PbTaSe$_2$ consists of stacked hexagonal TaSe$_2$ layers alternating with Pb monolayers. In the TaSe$_2$ layer, the Ta atom is located
offset from the inversion center. Electronic structure calculations reveal a single Dirac cone in the Brillouin zone which is gapped by
$\sim$0.8 eV due to large inherent SOC, in analogy to some topologically nontrivial materials \cite{Cava14}. On the other hand, the broken
inversion symmetry of the crystal structure along with the large SOC manifests itself as a large band splitting of order of several tenths
of an eV. All these features, namely, the lack of inversion symmetry and the strong SOC, appear to fulfill the necessary ingredients for
realizing a parity-mixed superconducting state and topological superconductivity. Nevertheless, the superconducting and normal state
properties of this material, which may provide valuable clues to the pairing mechanism, and potentially identify possible topological
states, have not yet been well characterized.

With this aim in mind, we use electric and thermal transport as a probe to study the superconducting and normal state properties of
PbTaSe$_2$. The upper critical field of PbTaSe$_2$ was determined by resistivity measurements down to 0.5 K \cite{Niu13,Xu14}, and revealed
a prominent upward curvature, regardless of the criteria used to determine $H_{c2}$. Detailed modelling of the $H_{c2}$ data further
suggests two-band superconductivity in PbTaSe$_2$. The thermoelectric power (TEP) shows the nonlinearity in $T$ and the dominance of
negative charge carriers. Remarkably, our pressure study demonstrates a clear V-shaped pressure dependence of $T_c$, similar to that
recently observed in AFe$_2$As$_2$ (A=K, Rb, Cs) \cite{Tafti13,Tafti14,Tafti15}. The origin of this V-shaped diagram in PbTaSe$_2$ is
possibly due to a Lifshitz transition under pressure, in contrast to the change of pairing symmetry in AFe$_2$As$_2$ (A=K, Rb, Cs). Our
findings may impose important constraints on any theory of exotic superconductivity in this system and provide some useful indications for
the presence or absence of topological states.

\section{Experiment}

Polycrystalline samples of PbTaSe$_2$ were prepared by a solid state reaction in evacuated quartz tubes, following similar procedures as
given in Ref. \cite{Cava14}. Good crystallinity of the samples was identified by a x-ray powder diffractometer. Traces of an impurity phase
of \textit{nonsuperconducting} tantalum oxide, were still detectable. Resistivity was measured by a standard four-probe lock-in technique
in a Quantum Design PPMS equipped with a 9 Tesla magnet. For the TEP measurements, a modified steady-state method was used in which a
temperature gradient, measured using a constantan-chromel differential thermocouple, was set up across the sample via a chip heater
attached to one end of the sample \cite{Xu12,Wakeham11,Bangura13}. The thermopower voltage was read out by a nanovoltmeter K2182A from
Keithley Instruments. For the hydrostatic pressure measurements, a commercial piston-type cell from Quantum Design was used, and Daphne
7373 oil was applied as the pressure transmission media. For the resistivity curves under pressure, the same contacts were used throughout
the measurements such that the geometric errors in the contact size were identical for different pressures.

\section{Results And Discussion}

Figure 1(a) presents the zero-field resistivity of PbTaSe$_2$ sample from 350 K down to 2 K. Zero resistivity of the superconducting
transition is clearly seen, with the low-$T$ transition expanded in Fig. 1(b). We also note that our transition is slightly broader and the
residual resisitivity ratio (RRR) is somewhat smaller than that reported in Ref. \cite{Cava14}, mostly because of the higher disorder level
or slight off-stoichiometry of Se in our samples, considering the high vapor pressure of Se in the reaction. The high-$T$ resistivity rises
linearly with temperature above $\sim$150 K. This is consistent with its Debye temperature $\Theta_D$=143 K estimated from the heat
capacity measurement, indicating the dominance of electron-phonon scattering at this $T$-range \cite{Ashcroftbook}. The superconducting
transition at various magnetic fields was studied by the fixed-field temperature sweeps down to $^3$He temperatures, presented in Fig.
1(c). In this study, four different criteria have been used to extract the critical temperature at each magnetic field, as exemplified in
Fig. 1(b). These criteria include: The onset $T_c^{onset}$, 90\% of the normal state $\rho_n$ $T_c^{90\%}$, middle point $T_c^{50\%}$ and
the zero resistance $T_c^{zero}$.

The resultant $H_{c2}$-$T$ diagram from these four criteria is cumulatively presented in Fig. 2(a). This diagram features, unlike most of
superconductors whose $H_{c2}$ has a negative/downward curvature, the evident upward curvatures for \textit{all} four lines. Note this
positive curvature is present for all $T$ below $T_c$ and it becomes very steep for the $T_c^{onset}$ line at low $T$. This overall upward
curvature can not be explained by the Ginzburg-Landau theory, nor by the one-band Werthamer-Helfand-Hohenberg (WHH) model \cite{WHH66}. In
Ref. \cite{Cava14}, the formula $H_{c2}(T)$=$H_{c2}$(0)(1-$t$$^{3/2}$)$^{3/2}$ was applied to fit the data between 2.4 K to $T_c$. However,
we show in Fig. 2(b) that this formula actually deviates from the experimental data below 1.8 K. Instead, it is possible to capture the
whole data set with the two-band theory, as the upward $H_{c2}$ has also been observed in some two-gap superconductors, e.g.,
LaFeAsO$_{0.89}$F$_{0.11}$ \cite{Hunte08}, MgB$_2$ \cite{MgB2Lima01} and Ca$_{10}$(Pt$_4$As$_8$)((Fe$_{1-x}$Pt$_x$)$_2$As$_2$)$_5$
\cite{Ni12}. To analyze our experimental data on a quantitative footing, we now fit the $H_{c2}$ curve with the two-band theory outlined in
Ref. \cite{Gurevich03}. The equation of $H_{c2}(T)$ for a two-band superconductor is given by:

\begin{eqnarray}
a_0[lnt+U(h)][lnt+U(\eta h)]+a_1[lnt+U(h)]\nonumber\\
+a_2[lnt+U(\eta h)]=0,\label{eqn:one}
\end{eqnarray}

\noindent where $a_0$=2($\lambda_{11}$$\lambda_{22}$-$\lambda_{12}$$\lambda_{21}$)/$\lambda_0$,
$a_1$=1+($\lambda_{11}$-$\lambda_{22}$)/$\lambda_0$, $a_2$=1-($\lambda_{11}$-$\lambda_{22}$)/$\lambda_0$,
$\lambda_0$=$\sqrt{(\lambda_{11}-\lambda_{22})^2+4\lambda_{12}\lambda_{21}}$, $h$=$\frac{H_{c2}D_1}{2\phi_0T}$, $\eta=\frac{D_2}{D_1}$ and
$U(x)=\psi(x+\frac{1}{2})-\psi(\frac{1}{2})$. $\psi(x)$ here is the digamma function. $\lambda_{11}$, $\lambda_{22}$ denote the intraband
coupling constants, and $\lambda_{12}$, $\lambda_{21}$ are the interband coupling constants. $D_1$ and $D_2$ are the diffusivity of each
band. The small $\eta$ may imply a much stronger scattering on one of the bands. Hence, there are totally six free parameters in the
fitting process, $\lambda_{11}$, $\lambda_{22}$, $\lambda_{12}$, $\lambda_{21}$, $D_1$ and $\eta$. Within this theoretical framework, we
fit our experimental data from the 90$\%$ of $\rho_n$ criterion, as shown in Fig. 2(b). This two-band theory can overall fit the
experimental data within the error bar and the resultant parameters are given in the figure. As can be seen, the intraband coupling
slightly dominates the interband coupling while $\eta$$\ll$1, indicating that electrons on one band are more scattered than on the other.
However, it can \textit{not} be concluded that the two-band superconductivity is wholly responsible for the upward curvature of $H_{c2}$ in
this material, in view of the number of free parameters in the fit. In fact, we also fit the data from 50$\%$ of $\rho_n$ criterion and
obtain $\lambda_{11}$=$\lambda_{22}$=0.5, $\lambda_{12}$=$\lambda_{21}$=0.4, $D_1$=0.000575 and $\eta$=0.071, also suggestive of the much
stronger scattering on one band. Although these $H_{c2}$ data can not exclusively rule out other possible scenarios, which will be
discussed later, they do indicate that two-band superconductivity appears to be a plausible mechanism in PbTaSe$_2$ \cite{footnote1}.

Let us discuss other possible scenarios for the upward curvature in $H_{c2}$. In the organic superconductor (TMTSF)$_2$PF$_6$, $H_{c2}$
also has a pronounced upward curvature and no sign of saturation was observed down to 0.1 K \cite{Chaikin97}. However, its $H_{c2}$ at
$T\rightarrow0$K exceeds the Pauli paramagnetic limit by a factor of 2-3. Therefore, the upward curvature was attributed to the signature
of spin-triplet pairing. In the case of PbTaSe$_2$, its $H_{c2}$ at 0.5 K is still much lower than its Pauli limit, suggesting that the
orbital effect dominates in limiting its $H_{c2}$. One theory accounting for the strong upward curvature observed in the cuprates
\cite{Mackenzie93,Osofsky93} was associated with the quantum critical point in these systems \cite{Varma96}. In our system, however, there
is no clear sign to show it is in proximity to the quantum critical regime.

The Seebeck coefficient or TEP reveals important aspects of charge conduction in a material \cite{Cohn12,Cohn14}. Figure 3 shows the
Seebeck coefficient $S(T)$ of PbTaSe$_2$ from 350 K down to 5 K. Evidently, its TEP, of modest magnitude, is negative over the entire
$T$-range studied, indicating the dominance of electron carriers. In a single-band metal, according to the Sommerfeld theory, TEP is given
by $S$=-$\frac{\pi^2}{2}$$\frac{k_B^2}{e}$$\frac{T}{E_F}$, where $E_F$ is the Fermi energy \cite{Ashcroftbook,Behnia09}. This leads to a
$T$-linear Seebeck coefficient. In reality, the situation may become complicated due to the presence of other excitations. In a metal with
both electrons and holes, the electron and hole partial Seebeck coefficients add together according to their weights in conduction
\cite{Cohn12,Cohn14}, i.e., $S$=($\frac{\sigma_h}{\sigma}$)$S_h$+($\frac{\sigma_e}{\sigma}$)$S_e$. $S_h$ and $S_e$ are the Sommerfeld
values (both linear in $T$) given above. Provided that electron and hole partial weights in conduction also change with $T$, a
nonlinear-in-$T$ Seebeck coefficient can naturally emerge. Indeed, from \textit{ ab initio }calculations \cite{Cava14}, the Fermi surface
of PbTaSe$_2$ consists of both electron and hole pockets. On the other hand, since the electron pocket dominates, we may apply the
single-band Sommerfeld equation quoted above to make a \textit{crude} estimate of the Fermi energy of this material. From Fig. 3, the
linear fit at low $T$ gives $E_F$$\sim$480 meV.

Shown in Fig. 4(a) is the pressure dependence of the resistivity of PbTaSe$_2$ up to $\sim$25 kbar. With increasing pressure, the
resistivity of the sample decreases systematically, while the pressure seems to have little bearing on its overall $T$ dependence. The
resistivity at room temperature changes by about 15$\%$ at 25 kbar. Interestingly, although the metallicity of the sample is enhanced by
the pressure, its $T_c$ shows a nonmonotonic variation, from an initial decrease with pressure to an increase once across a critical
pressure, as demonstrated in Fig. 4(b) and 4(c). Note that $T_c^{onset}$ of this sample is about 0.2 K lower than that presented in Fig.
1(b). This discrepancy may arise from the sample off-stoichiometry even if they were harvested from the same batch. Utilizing the same
criteria as in Fig. 1(b), we plot the pressure dependence of $T_c$ in Fig. 5(a). Remarkably, regardless of the criteria used, $T_c$ is
firstly reduced by the pressure up to $P_c$$\sim$5-10 kbar, after which it starts to reverse to higher $T_c$, similar to the effects of
hydrostatic pressure on iron-based superconductors AFe$_2$As$_2$ (A=K, Rb, Cs) \cite{Tafti13,Tafti14,Tafti15}.

The origin of this $T_c$ reversal under pressure has been extensively discussed by F. Tafti \textit{et al}. in their works
\cite{Tafti13,Tafti14,Tafti15}. This includes a change of the pairing symmetry or a Lifshitz transition across $P_c$, i.e., an abrupt
change of Fermi surface topology. In AFe$_2$As$_2$ (A=K, Rb, Cs), they argued that the V-shaped pressure dependence of $T_c$ was the result
of a change in pairing symmetry from $d$-wave state below $P_c$ to an $s_\pm$ above $P_c$. A Lifshitz transition was ruled out in their
data, as there were no visible changes in either the Hall or resistivity data across $P_c$. Following the same procedure given in Ref.
\cite{Tafti15}, we plot the inelastic scattering, defined as $\rho$(T=20 K)-$\rho_0$ where $\rho_0$ is the residual resistivity at each
pressure, as a function of pressure in Fig. 5(b). We should mention that the choice of $T$=20 K is arbitrary but any other cuts above $T_c$
do not affect the results. The plot of $\rho$(20K) in the inset also shows a similar dependence. In contrast to AFe$_2$As$_2$ (A=K, Rb,
Cs), the inelastic scattering in PbTaSe$_2$ displays a sharp drop at a critical pressure somewhere between 5 kbar to 10 kbar (the shaded
area in the figure), where $T_c$ also shows a sharp inversion. This makes a compelling case that the $T_c$ reversal in this system is most
likely due to a Lifshitz transition around $P_c$ (5-10 kbar).

\section{Conclusion}

To summarize, we report a strong positive curvature in the $H_{c2}$-$T$ diagram of the NCS superconductor PbTaSe$_2$. Its Seebeck
coefficient was found to be negative in sign and to vary non-linearly with $T$. All these experimental data appear to be consistent with a
picture of two-band superconductivity in PbTaSe$_2$, while other possible mechanisms have also been discussed. Interestingly, the recent
theoretical proposal of the large Seebeck coefficient arising from topological Dirac fermions has not been seen in our samples
\cite{Zhang14}. The pressure dependence of the superconducting transition $T_c$ shows a clear 'V' shape, which most likely results from a
Lifshitz transition under pressure. Further studies are highly desired to explore the possible parity-mixed state by using ultralow-$T$
thermal conductivity and/or penetration depth measurements.

\section{Acknowledgement}

The authors would like to thank N. E. Hussey, C. M. J. Andrew, C. Lester, A. F. Bangura, Xin Lu, Zengwei Zhu, Xiaofeng Jin for valuable
discussions. This work was supported by the National Key Basic Research Program of China (Grant No. 2014CB648400) and by NSFC (Grant No.
11474080, 11104051, 11104053). X.X. would also like to acknowledge the auspices from the Distinguished Young Scientist Funds of Zhejiang
Province (LR14A040001).

\pagebreak[5]

\begin{figure}
\begin{center}
\vspace{-.2cm}
\includegraphics[width=12cm,keepaspectratio=true]{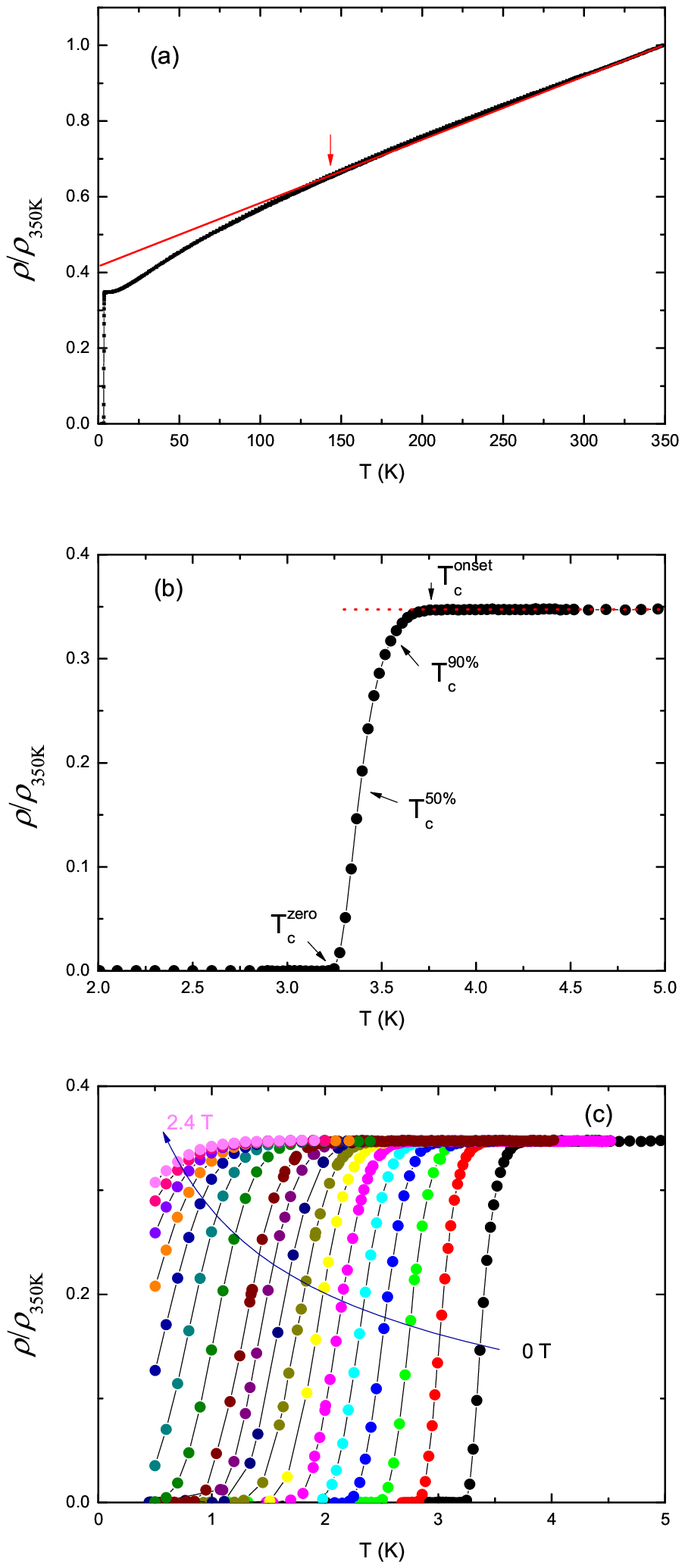}
\caption{(Color online) (a) Zero-field resistivity of PbTaSe$_2$ measured from 350 K down to 2 K. The data were renormalized to the 350 K
value. Resistivity shows a clear deviation from high-$T$ linearity (red straight line) below $\sim$150 K, marked by the arrow. (b) A
close-up view of the low-$T$ superconducting transition, illustrating the four criteria used to determine $T_c$ at each field. (c)
Superconducting transition under various magnetic fields.} \label{Fig1}
\end{center}
\end{figure}

\begin{figure}
\begin{center}
\vspace{-.2cm}
\includegraphics[width=12cm,keepaspectratio=true]{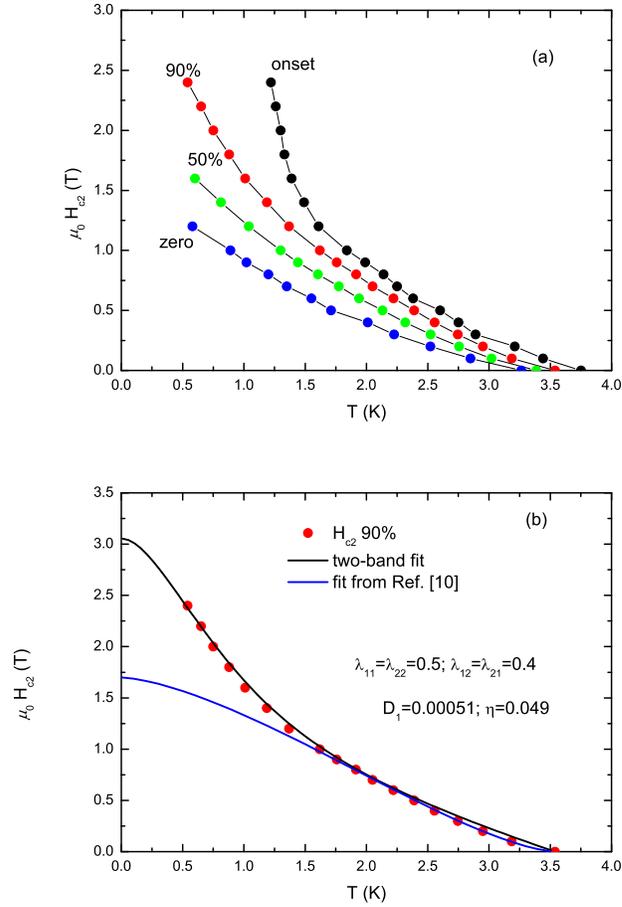}
\caption{(Color online) (a) Temperature dependence of the upper critical field $H_{c2}$ determined from Fig. 1(c) by using the criteria in
Fig. 1(b). (b) The experimental $H_{c2}$ data using the 90$\%$ of $\rho_n$ criterion, along with the two-gap fitting described in the text.
The blue thin line shows the failure of the fit to the equation $H_{c2}(T)$=$H_{c2}$(0)(1-$t$$^{3/2}$)$^{3/2}$, where $t$ is the reduced
temperature $T/T_c$, used in Ref. \cite{Cava14}. } \label{Fig2}
\end{center}
\end{figure}

\begin{figure}
\begin{center}
\vspace{-.2cm}
\includegraphics[width=15cm,keepaspectratio=true]{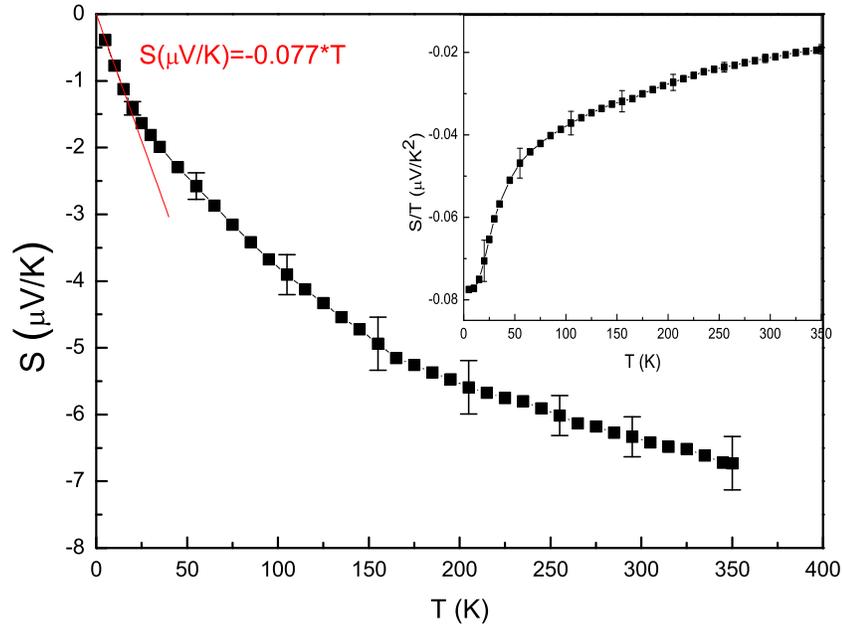}
\caption{(Color online) The temperature dependence of the Seebeck coefficient measured from 350 K down to 5 K. The typical error bars,
estimated from different runs with varying heater currents, were given at certain temperatures. The straight red line is the linear fit to
the low-$T$ data. The inset shows the plot of $S/T$ as a function of temperature. } \label{Fig3}
\end{center}
\end{figure}

\begin{figure}
\begin{center}
\vspace{-.2cm}
\includegraphics[width=15cm,keepaspectratio=true]{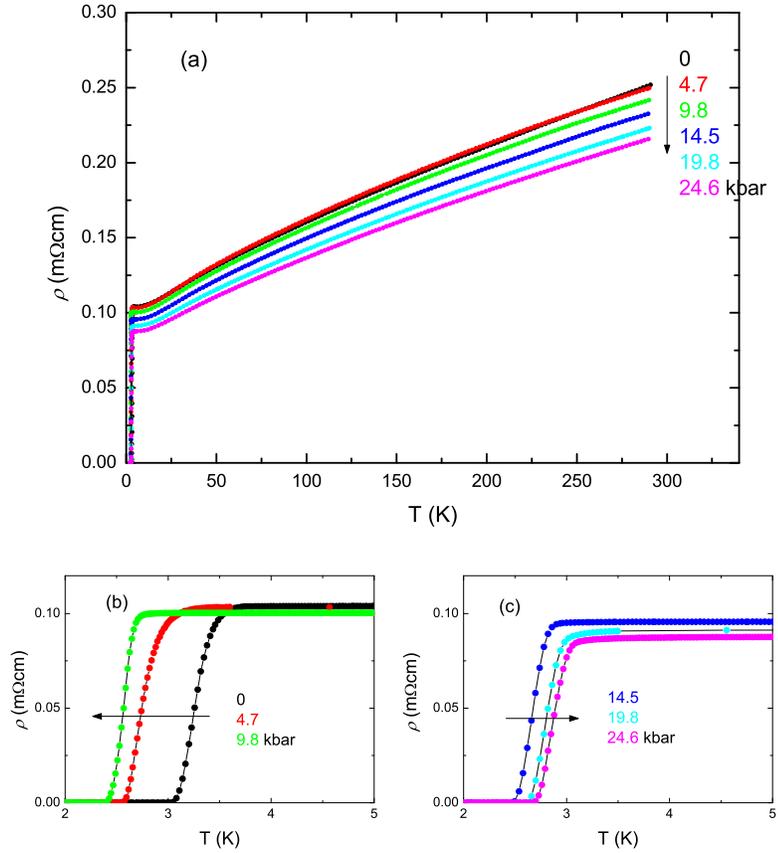}
\caption{(Color online) (a) shows the resistivity under several pressures, measured by the same electrical contacts. Panels (b) and (c)
expand the low-$T$ superconducting transitions under different pressure values. } \label{Fig4}
\end{center}
\end{figure}

\begin{figure}
\begin{center}
\vspace{-.2cm}
\includegraphics[width=15cm,keepaspectratio=true]{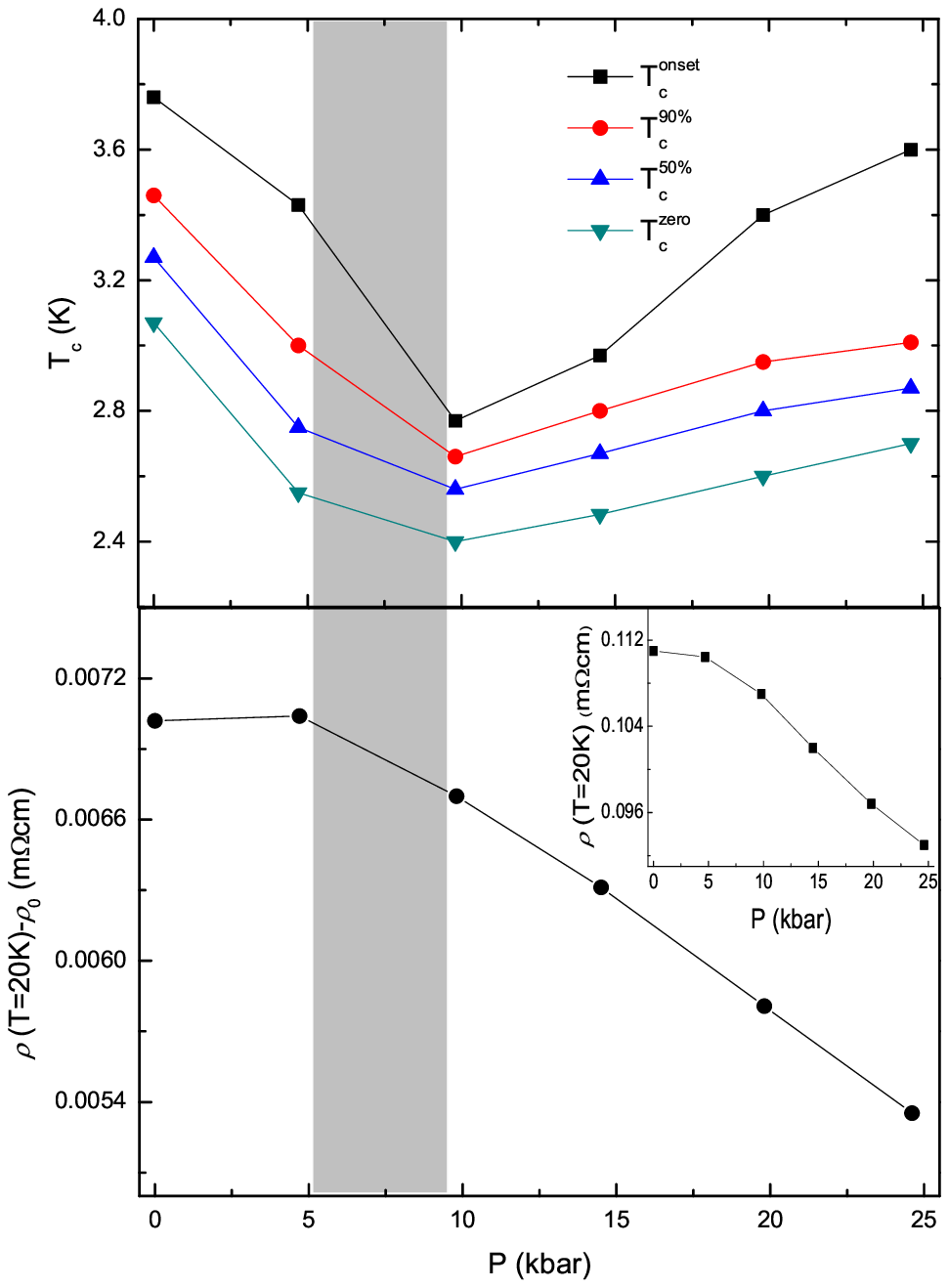}
\caption{(Color online) (a) shows $T_c$, estimated by four criteria used in Fig. 1(b), as a function of pressure. (b) The inelastic
scattering, defined as $\rho$(T=20 K)-$\rho_0$ where $\rho_0$ is the residual resistivity at each pressure, as a function of pressure. The
critical pressure where the Lifshitz transition occurs is somewhere in the shaded area. The inset shows $\rho$(T=20 K) versus pressure.}
\label{Fig5}
\end{center}
\end{figure}


\begin{thebibliography}{32}
\bibitem{footnote1} In principle, the Hall effect may be used to identify the multiband behaviors. However, the Hall signal in this material is
difficult to measure, possibly because of two reasons. First, in the Hall measurements, the sample needs to be thin in order to get large
Hall signals, however, our sample is very hard to be cleaved or polished to a thin plate. Second, in a two-carrier model, the Hall
coefficient at low fields is given by $R_H$=$\frac{\sigma_h^2 R_h +\sigma_e^2 R_e}{(\sigma_h +\sigma_e)^2}$, where $\sigma_{e(h)}$ and
$R_{e(h)}$ are the conductivity and Hall coefficient of $e(h)$ carriers, respectively (Note that $R_e$ and $R_h$ are opposite in sign). The
resultant Hall effect may become very small due to the substantial cancellation of hole and electron components.
\end{thebibliography}
\end{document}